\documentclass[
superscriptaddress,
 amsmath,amssymb,
 aps,
reprint
]{revtex4-2}

\usepackage[utf8]{inputenc}
\usepackage{color}
\usepackage{graphicx}
\usepackage{dcolumn}
\usepackage{bm}
\usepackage{epstopdf}
\usepackage{gensymb}
\usepackage{hyperref}

\usepackage{float}

\begin{document}


\title{How transparent is graphene?\\
A surface science perspective on remote epitaxy}

\author{Zachary LaDuca}
\affiliation{Materials Science and Engineering, University of Wisconsin-Madison, Madison, WI 53706, United States of America}

\author{Anshu Sirohi}
\affiliation{Materials Science and Engineering, University of Wisconsin-Madison, Madison, WI 53706, United States of America}

\author{Quinn Campbell}
\affiliation{Sandia National Laboratories, Albuquerque, NM 87123, United States of America}

\author{Jason K. Kawasaki} \email{Author to whom correspondence should be addressed: Jason Kawasaki, jkawasaki@wisc.edu}
\affiliation{Materials Science and Engineering, University of Wisconsin-Madison, Madison, WI 53706, United States of America}

\date{\today}

\begin{abstract}

Remote epitaxy is the synthesis of a single crystalline film on a graphene-covered substrate, where the film adopts epitaxial registry to the substrate as if the graphene is transparent. Despite many exciting applications for flexible electronics, strain engineering, and heterogeneous integration, an understanding of the fundamental synthesis mechanisms remains elusive. Here we offer a perspective on the synthesis mechanisms, focusing on the foundational assumption of graphene transparency. We identify challenges for quantifying the strength of the remote substrate potential that permeates through graphene, and propose Fourier and beating analysis as a bias-free method for decomposing the lattice potential contributions from the substrate, from graphene, and from surface reconstructions, each at different frequencies. We highlight the importance of graphene-induced reconstructions on epitaxial templating, drawing comparison to moiré epitaxy. We highlight the role of the remote potential in tuning surface diffusion and adatom kinetics on graphene, which are crucial for navigating the competition between remote epitaxy and defect-seeded mechanisms like pinhole epitaxy. In light of this weak remote potential, we re-evaluate the current state-of-the-art experimental evidence, highlighting why it remains challenging to experimentally validate a ``remote'' epitaxy mechanism that cannot be explained by alternatives, such as pinhole-seeded epitaxy or serial van der Waals epitaxy. We end with one experimental example that, to out knowledge, cannot be explained by competing mechanisms: a different long-range epitaxial relationship for GdPtSb films grown on graphene/sapphire, compared to direct epitaxy on sapphire. We suggest for future experiments that directly measure the remote potential and impact of tuneable growth kinetics.

\end{abstract}

\maketitle

\section{Introduction}

Remote epitaxy\cite{kim2017remote} is an approach for synthesizing single crystalline films that are semi-decoupled from their substrates, to circumvent the lattice \cite{hull1992misfit, fitzgerald1991totally} and chemical \cite{palmstrom1995epitaxy} mismatch challenges of traditional epitaxy. In this approach, thin films are grown on a graphene-covered single crystalline substrate. Under the right conditions, a single crystalline film can be produced which has epitaxial registry to the substrate rather than to graphene. To date, this concept has been applied to the synthesis of III–V and III–N semiconductors \cite{kim2017remote, HENKSMEIER2022InGaAs, Bae2020GaPGraphene,jiang2019carrier, liu2022atomic}, transition metal oxides\cite{kim2022SciAd,yoon2022freestanding}, halides\cite{yuan2025remote, jiang2019carrier}, and intermetallic compounds\cite{laduca2024cold,du2023strain,du2021epitaxy, du2022controlling}. Applications of these heterostructures include membrane exfoliation for flexible electronics \cite{kim2022chip, jeong2020selective}, discovery of new properties via strain engineering in membranes \cite{du2021epitaxy, laduca2024cold,dai2022highly, kum2020heterogeneous, laduca2023control}, lattice mismatched heteroepitaxy with reduced dislocation densities \cite{Bae2020GaPGraphene, jiang2019carrier, liu2022atomic}, integration of dissimilar materials \cite{yuan2025remote, dai2022highly}, e.g., for stacking multi color (bandgap) micro-LED arrays \cite{Shin2023_stackedLED}, blocking diffusion across film/substrate interfaces\cite{strohbeen2021quantifying}, and reuse of expensive substrates\cite{Kim2023_highthroughput}. Several reviews have focused on applications \cite{roh2023applications, du2023strain, park2024remote}, the processing steps required for making, transferring, and performing epitaxy on top of graphene-covered surfaces \cite{kim2022remote, park2024remote, ji2023understanding}, and comparison with different types of epitaxial growth modes \cite{ryu2022two, liu2023two}.

However, the synthesis mechanisms of remote epitaxy remain elusive. The foundational assumption of remote epitaxy is that graphene is transparent and weakly interacting. That is, graphene allows the lattice potential of the substrate to pass through while contributing a relatively weak lattice potential of its own, and the graphene does not distort the underlying substrate atomic positions or vice versa. This degree of transparency is difficult to quantify due to challenges in preparing clean graphene/substrate interfaces \cite{manzo2022pinhole, kim2021role, kim2021impact, yoon2022freestanding, du2022controlling}. The strength of the remote potential is typically inferred via post film growth observables like epitaxial alignment of the substrate or ability to exfoliate a membrane \cite{kong2018polarity,kim2017remote} or its apparent wetting transparency \cite{rafiee2012wetting, shih2012breakdown}, but has not been directly measured \cite{ke1999quantity,allain2017color,bauer1994low}. Additionally, there are many examples where graphene induces strong distortions in an underlying substrate. One example is ``buffer'' graphene that forms on Si-terminated SiC (0001) \cite{GrSiC_Bonding_XRSW}. This buffer graphene is highly buckled due to partial covalent bonding to the SiC substrate, which results in dangling bonds out-of-plane and a long-range reconstruction in-plane with $(6 \sqrt{3} \times 6 \sqrt{3})R30 \degree$ periodicity. Buffer graphene also has a semiconducting gap \cite{n2017band}, suggesting less free carrier screening of the remote potential through buffer graphene than through pristine semimetallic graphene. These factors suggest that graphene-induced surface reconstructions are important to the mechanisms of remote epitaxy, beyond the simplest limit of graphene transparency.

Here we highlight some of the outstanding questions regarding graphene transparency, their impacts on thermodynamics and kinetics of remote epitaxy, and why it remains an outstanding challenge to definitively demonstrate a remote epitaxy mechanism that cannot be explained by alternatives like pinholes \cite{manzo2022pinhole} or serial van der Waals epitaxy:

\textbf{Section \ref{sec:potential}: }How strong is the remote lattice potential? What is the nature of the remote bonding interaction and what kind of potential is the correct descriptor? How can we analyze complicated potentials that convolve the substrate lattice with the graphene lattice?

\textbf{Section \ref{sec:reconstruction}: } How does graphene interact with an underlying substrate? What are the impacts of changes in graphene bonding, local atomic registry between graphene and substrate, and surface reconstructions, on epitaxial growth on these surfaces?  

\textbf{Section \ref{sec:kinetics}: }How does the remote potential modify adatom kinetics on graphene? How does kinetics control remote epitaxy versus defect-seeded
mechanisms?

\textbf{Section \ref{sec:experiment}:} Given the weak remote substrate potential, how can we explain existing experiments on ``remote'' epitaxy?

\textbf{Section \ref{sec:outlook}:} Outlook and call for new experiments and theory.

\section{How strong is the remote lattice potential?} \label{sec:potential}

\begin{figure*}
    \centering
    \includegraphics[width=1\linewidth]{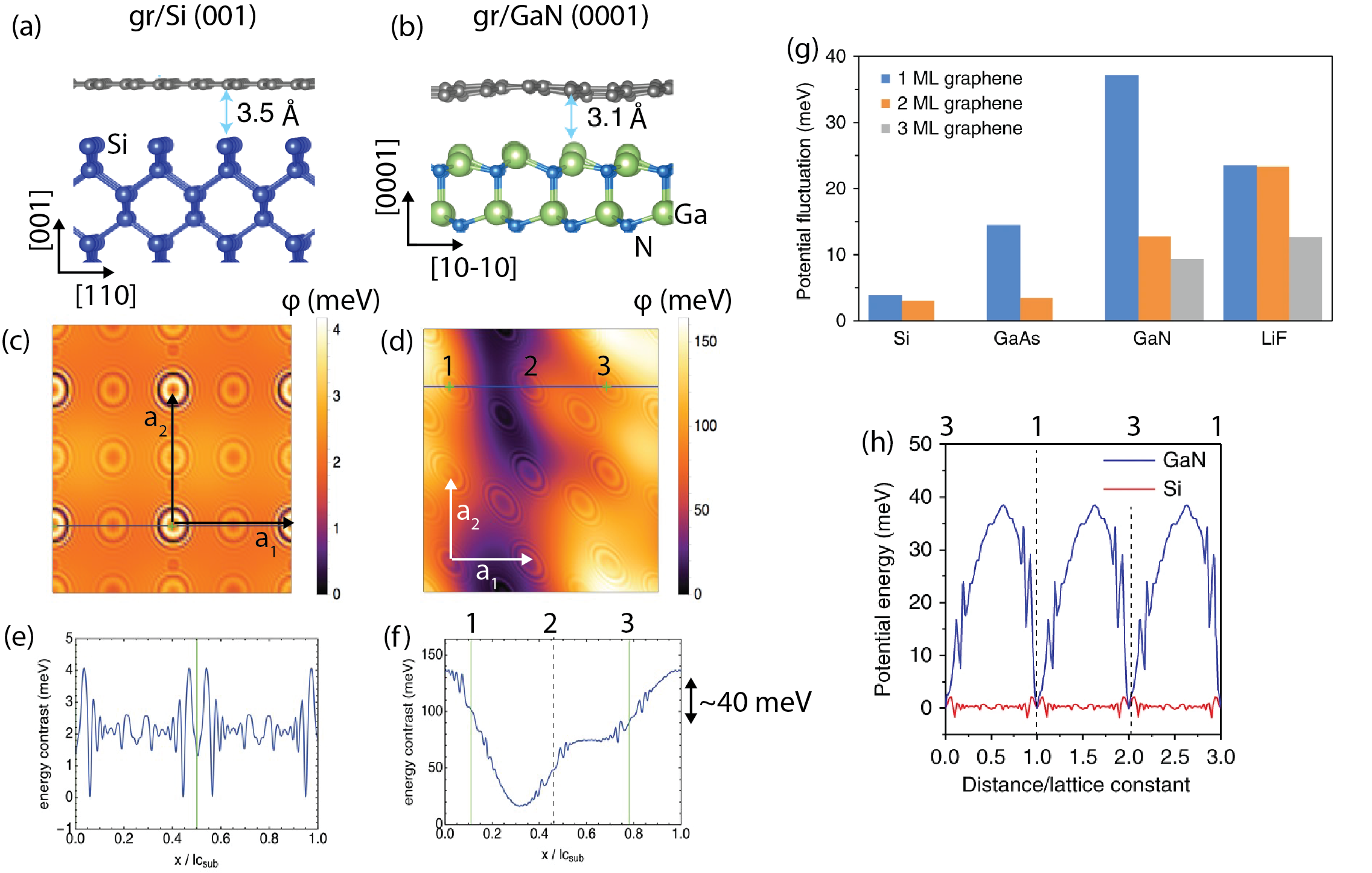}
    \caption{\textbf{Challenges for interpreting electrostatic potential maps of graphene/substrate heterostructures.} (a,b) Crystal structures of graphene/Si (001) and graphene/GaN (0001), with fixed in-plane positions and allowed out-of-plane relaxation. (c,d) Electrostatic potential maps. The potential fluctuations of the graphene itself have been subtracted away. Atom position labels 1, 2, 3 and crystallographic axes are added for clarity. Note that (d) is sheared from the hexagonal symmetry of GaN (0001). (e,f) Line cuts of the electrostatic potentials. (g) Summary of extracted $\Delta \varphi$ for different graphene covered substrates. (h) Periodic profiles of the selected regions from (e) and (f). For graphene/Si it is the region $x=0$ to $x=0.5$ in (e). For graphene/GaN it is the region $x\approx 0.8$ to $x\approx 0.1$ (atom 3 to atom 1) in (d).
    Adapted from Kong \textit{et. al.}, ``Polarity governs atomic interaction through two-dimensional materials,'' Nature Materials 17, 999–1004 (2018) Springer Nature \cite{kong2018polarity}. Reproduced with permission from Springer Nature.}
    \label{polarity}
\end{figure*}

We first address the strength of the remote lattice potential permeation of the substrate. For simplicity we ignore reconstructions here, and revisit the reconstructions in Section \ref{sec:reconstruction}. Quantifying this potential is challenging since the nature of the film-graphene-substrate interaction remains unclear. In conventional epitaxy and van der Waals (vdW) epitaxy, where the characteristic spacing between film and substrate is $2-3$ \AA, the primary interactions are covalent bonding and vdW interactions (fluctuating dipoles) between atoms. But for remote epitaxy, where film and substrate are separated by $\sim 6$ \AA\ due to insertion of graphene, the nature of this interaction is unclear. One might expect it is some combination of covalent and vdW bonding interaction plus longer range electrostatics (if the adatoms are ionized, e.g. by charge transfer to or from graphene, as often occurs for alkali metals in graphite intercalation compounds \cite{dresselhaus1981intercalation}). Note, however, that electrostatics are less relevant for neutral adatoms. Another possibility is an extended molecular orbital that includes states from the substrate that hybridize with graphene \cite{wang2023modulation}.

To date, the field has relied on several proxies for the remote lattice potential. In the pioneering work on remote epitaxy, Kim \textit{et. al.} \cite{kim2017remote} computed the charge density as a function of distance for GaAs slabs that are separated by a vacuum gap, using density functional theory (DFT). They found that for slabs separated by up to 9 \AA, there is a finite charge density in the middle of the vacuum gap. This calculation provided the first theoretical suggestion that the lattice potential of the substrate may extend significantly beyond the $2-3$ \AA\ typical equilibrium spacing between atoms. It should be noted, however, that within the DFT surface science community, the emergence of charge within the vacuum region is considered a sign that the vacuum region is not large enough to create an ``isolated'' surface. While the calculations in question are trying to demonstrate that interaction between a substrate and film can take place across larger vacuum distances, this ignores the periodic nature of the structure. \textit{I.e.}, the DFT calculation represents an infinite GaAs material with repeating vacuum regions in the middle. The GaAs in the simulation is then playing the role of both the substrate and the film. It is not clear from these calculations with a single GaAs representing both the film and substrate that charge density interaction through the vacuum would persist if the film and substrate were truly isolated. A DFT slab calculation of this system would include separate GaAs atoms for the substrate and the film, both of which are separated from each other by substantial vacuum and, ideally, a dipole correction \cite{bengtsson1999dipole} such that the top of the film would not interact with the bottom of the substrate.

More importantly, many calculations of the remote potential (and its proxies) do not include free carrier screening from graphene, which is expected to be significant since graphene is a semimetal. As an order of magnitude estimate, the Thomas-Fermi screening length for a free electron gas with density $n_{3D} = 10^{20}$ cm$^{-3}$ is $\lambda_{TF}=1.7$ \AA. Assuming an effective graphene thickness of $\Delta z = 3$ \AA, this suggests that the field transmission through monolayer graphene is $T_s = exp(-\Delta z / \lambda_{TF}) = 17 \%$ and the transmission through bilayer graphene is $exp(-2\Delta z / \lambda_{TF}) = 3 \%$ \cite{kawasaki2025model}. Any estimate of the remote potential permeation through graphene would need to include screening.

The current standard is to use the electrostatic potential as a proxy for the remote interaction potential (Fig. \ref{polarity}). 
Generally the electrostatic potential is computed using DFT slabs. Some calculations include the graphene and then subtract it away during post processing \cite{kong2018polarity, qiao2021graphene}, while others ignore graphene entirely \cite{jiang2019carrier}. A common simplification is to ignore lateral reconstructions and fix the atoms to the bulk-like $(x,y)$ positions, since reconstructions would require much larger supercells. These calculations often allow for out-of-plane $z$ relaxations.

The electrostatic potential analysis was first introduced by Kong \textit{et. al.} \cite{kong2018polarity}, who concluded that the magnitude of lateral electrostatic potential fluctuations scales directly with the degree of bond polarity in the underlying substrate (Fig. \ref{polarity}(g)). Note that in these electrostatic potential maps (Fig. \ref{polarity}(c,d)), the lattice potential of graphene has already been subtracted away \cite{kong2018polarity}, making it difficult to compare the relative magnitude of the remote potential strength versus the graphene potential strength. This is important because, as we will discuss later, if the remote substrate potential and graphene potential are of similar magnitude, then epitaxy to graphene can compete with remote epitaxy to the substrate.
The bond polarity prediction was supported by experiments showing that for substrates with more polar bonding, single crystalline epitaxy was observed on a greater number of graphene layers. This provided a major guidance for the field: bond polarity controls remote epitaxy.

There are two major challenges for interpreting how the electrostatic potential guides epitaxy. First, although electrostatic (Coulomb) interactions describe the electronic workfunction and band alignments, it is less clear that electrostatics would control the pairwise attractive and repulsive interactions between atoms. The use of this potential assumes that the film adatoms are ionized. It is an open question these species are indeed ionized (and to what degree?), or whether other types of interactions, e.g. van der Waals interactions or covalent hybridization, may be more relevant to epitaxy and bonding. Kong \textit{et. al.} introduced an empirical scale factor $\gamma$ that describes the assumed degree of ionization. Then the modified electrostatic potential fluctuation is 
\begin{equation}
    \Delta \varphi = \gamma (\varphi_{max} - \varphi_{min}).
\end{equation}
The challenge with this scale factor is that the particular values of $\gamma$ are not constrained by measurement or theory. Ref. \cite{kong2018polarity} used values of $\gamma = 1$ for Si, GaAs, and GaN, and a value of $\gamma = 2$ for LiF. However, because the electronegativity and ionization energies vary signification for Si, GaAs, and GaN, it is unclear that $\gamma$ should equal 1 for the adatoms in all three materials. Further experiments and theory are required to constrain $\gamma$ and enable comparisons of $\Delta \varphi$ across different materials, with potentially different bonding character.

A second challenge is that the raw potential maps contain more than one periodicity: the graphene lattice, the substrate lattice, and a longer-range graphene-induced reconstruction, such that 
\begin{equation}
    \varphi_{total} =  \varphi_{gr} + T_s (\varphi_{sub} + \varphi_{rec})
\end{equation}
where $T_s$ is the free carrier screening coefficient through graphene \cite{kawasaki2025model}. To distinguish $\varphi_{sub}$ of the remote substrate, Ref. \cite{kong2018polarity} used the following subtraction and cropping procedure: First, they subtract the short period potential for isolated graphene ($\varphi_{gr}$), resulting in the maps of Fig. \ref{polarity}(c,d). Next they crop the longer range fluctuation of the relaxed slab ($\varphi_{rec}$): for the profile above graphene/GaN (0001) (Fig. \ref{polarity}(d,f)), they focus on the line cut from atom 3 to atom 1, crop away the rest of the slab, then subtract a linear background. The energy difference along the path from atom 3 to atom 1 is then used as $\varphi_{sub}$ for the substrate, and this selected profile is periodically repeated in the line cut shown in Fig. \ref{polarity}(h). 

\begin{figure}
    \centering
    \includegraphics[width=1\linewidth]{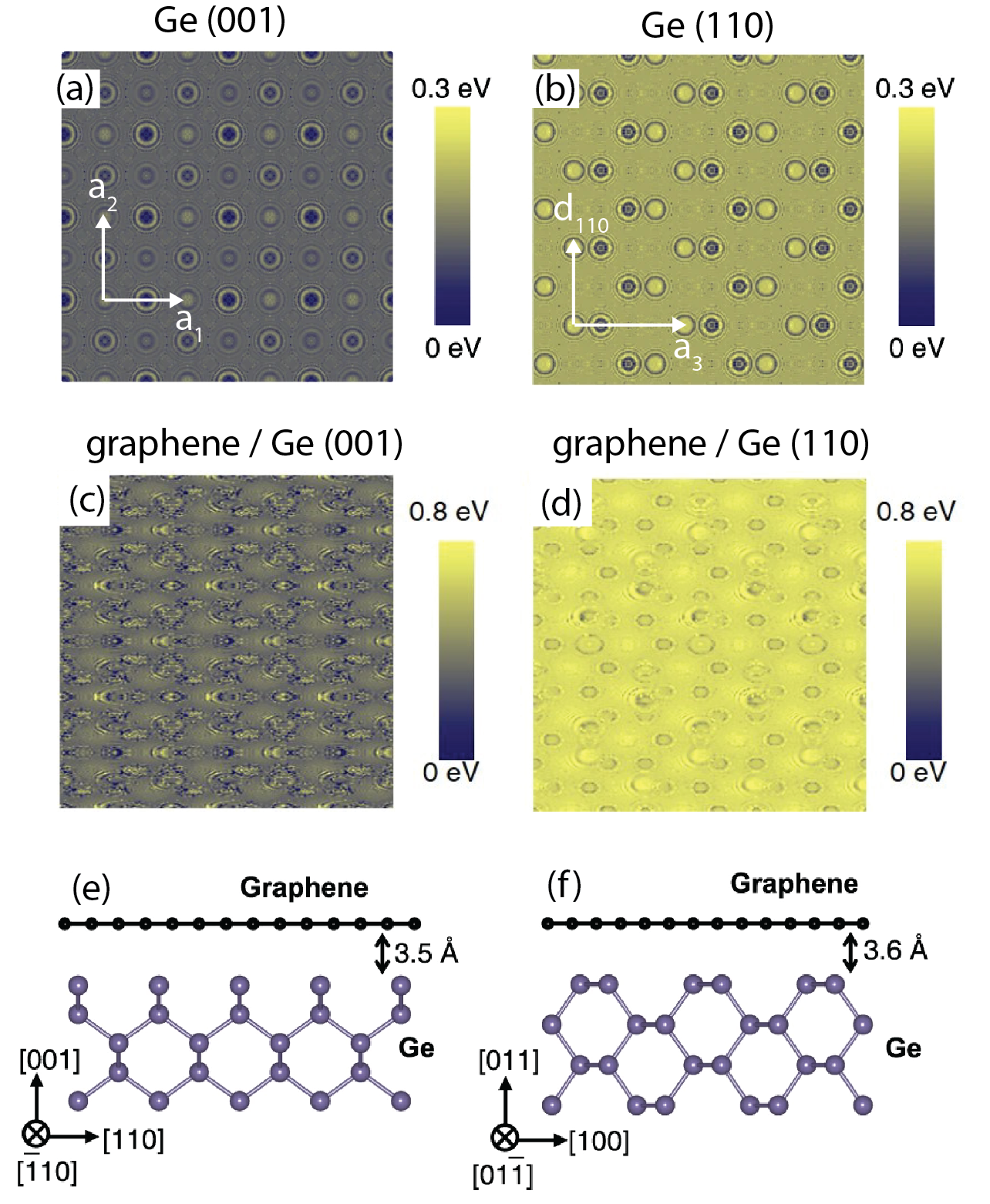}
    \caption{\textbf{Electrostatic potential calculations highlighting the interference from the graphene and substrate potentials.} Lattice vectors are added for clarity. Adapted from Dai \textit{et. al.} ``Highly heterogeneous epitaxy of flexoelectric BaTiO$_{3-\delta}$ membrane on Ge,'' Nature Commun., 13, 2990 (2022), Springer Nature \cite{dai2022highly}, under Creative Commons CC BY license.}
    \label{interference}
\end{figure}

The challenge with the background and cropping procedure is that it is difficult to determine which region of the line cut is most representative of $\varphi_{sub}$. Could it instead be defined as the region from atom 1 to atom 2, which has similar magnitude but opposite sign? Or could it be defined by the flat region from atom 2 to atom 3? Or should one consider the entire length of the supercell? The choice of region changes the answer from $\Delta \varphi_{sub} \approx 40$ meV (atom 3 to 1 or atom 1 to 2), to 80 meV (full range), to just a few meV (atom 2 to 3). Moreover, the immediate subtraction of $\varphi_{gr}$ makes it difficult to assess the relative magnitude of the remote potential fluctuation from the graphene potential fluctuation. For remote epitaxy to dominate, presumably $\Delta \varphi_{sub}$ would need to be much larger than $\Delta \varphi_{gr}$.

Calculations by Dai \textit{et. al.} \cite{dai2022highly} suggest that the graphene and remote substrate potentials have similar magnitude, resulting in complicated interference patterns (Fig. \ref{interference}). The electrostatic potentials above the bare Ge (001) and (110) surfaces match the lattice periodicity of Ge in those orientations (Fig. \ref{interference}(a,b)). The potential maps above graphene/Ge (001) and graphene/Ge (110) reveal more complicated patterns where the underlying periodicity square lattice periodicity of Ge (001) and rectangular lattice of Ge (110) are still observed, but convolved with the hexagonal graphene lattice (Fig. \ref{interference}(c,d)). These interference patterns suggest that the electrostatic potential fluctuations from Ge and graphene have similar magnitudes.

\begin{figure}[t]
    \centering
    \includegraphics[width=0.8\linewidth]{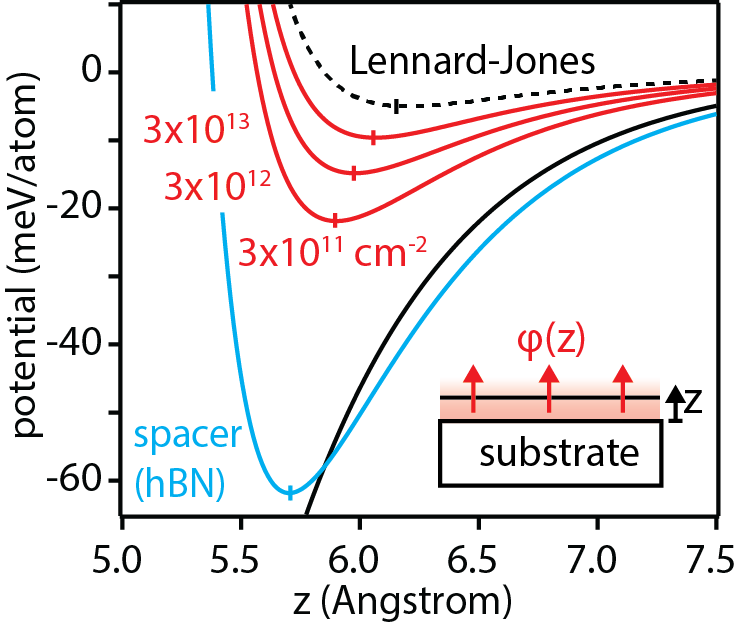}
    \caption{\textbf{Screened Morse model for the remote potential above graphene/GaAs (001).} Blue curve is the potential through an insulating hBN spacer, in which there is no screening. Red curves are the screened potential through graphene for different graphene carrier densities. Dotted black curve is the Lennard-Jones potential of graphene. Solid black curve is the Morse potential for a bare GaAs substrate.
    Adapted from Kawasaki and Campbell, ``An analytical model for the remote epitaxial potential,'' arXiv:2507.09913 (2025) \cite{kawasaki2025model}.}
    \label{fig:morse}
\end{figure}

A recent analytical model by Kawasaki and Campbell further suggests that once free carrier screening from graphene is included, the remote substrate potential and the graphene potential have similar magnitude \cite{kawasaki2025model} (Fig. \ref{fig:morse}), and thus interference with graphene cannot be ignored. This model starts from a Morse pair interaction potential (rather than electrostatics), which includes covalent and van der Waals interactions between film and substrate, plus a van der Waals spacer layer modeled by a Lennard-Jones potential. The inclusion of free carrier screening for typical graphene carrier densities significantly attenuates the remote potential, such that for a typical graphene densities of order $10^{12}$ cm$^{-2}$, the screened potential above graphene/GaAs is of order 10 meV (black curves), similar to the Lennard-Jones potential of graphene itself (dotted line). In the limit where the carrier density goes to zero, which is analogous to replacing graphene with insulating hBN (blue curve), the screened potential has magnitude $\sim 60$ meV, still much smaller than the 2.1 eV covalent bond strength for direct Ga-As bonding.
The advantages of this approach are that the Morse potential describes a pairwise bonding interaction, rather than the purely electrostatic interaction computed by DFT, and is defined by simple, interpretable parameters that are well benchmarked by experiments and theory.

How can one analyze these interference patterns to isolated the contributions from graphene and the substrate, in a bias-free way? We propose beating and Fourier analysis. Fig. \ref{fig:reconstruction}(a-c) shows the essentials of this analysis, using a 1D model for the graphene and substrate lattices as sine waves with frequency $Q_{gr}$ and $Q_{sub}$. The sum of these wave produces an interference pattern with beat frequency $(Q_{sub}-Q_{gr})/2$ (dotted line). In reciprocal space, the Fourier transform displays peaks at $Q_{gr}$ and $Q_{sub}$, allowing a bias-free analysis of the relative strengths of the graphene versus substrate potentials. Notably, the beat frequency does not appear in the Fourier analysis because it is not an independent frequency, it merely arises from interference between the two waves. These complementary real space and reciprocal space techniques may help to more rigorously define the different contributions to the potential fluctuation $\Delta \varphi$. 

\section{To what extent does graphene interact with a substrate or with a film?} \label{sec:reconstruction}

\begin{figure*}[t]
    \centering
    \includegraphics[width=0.9\linewidth]{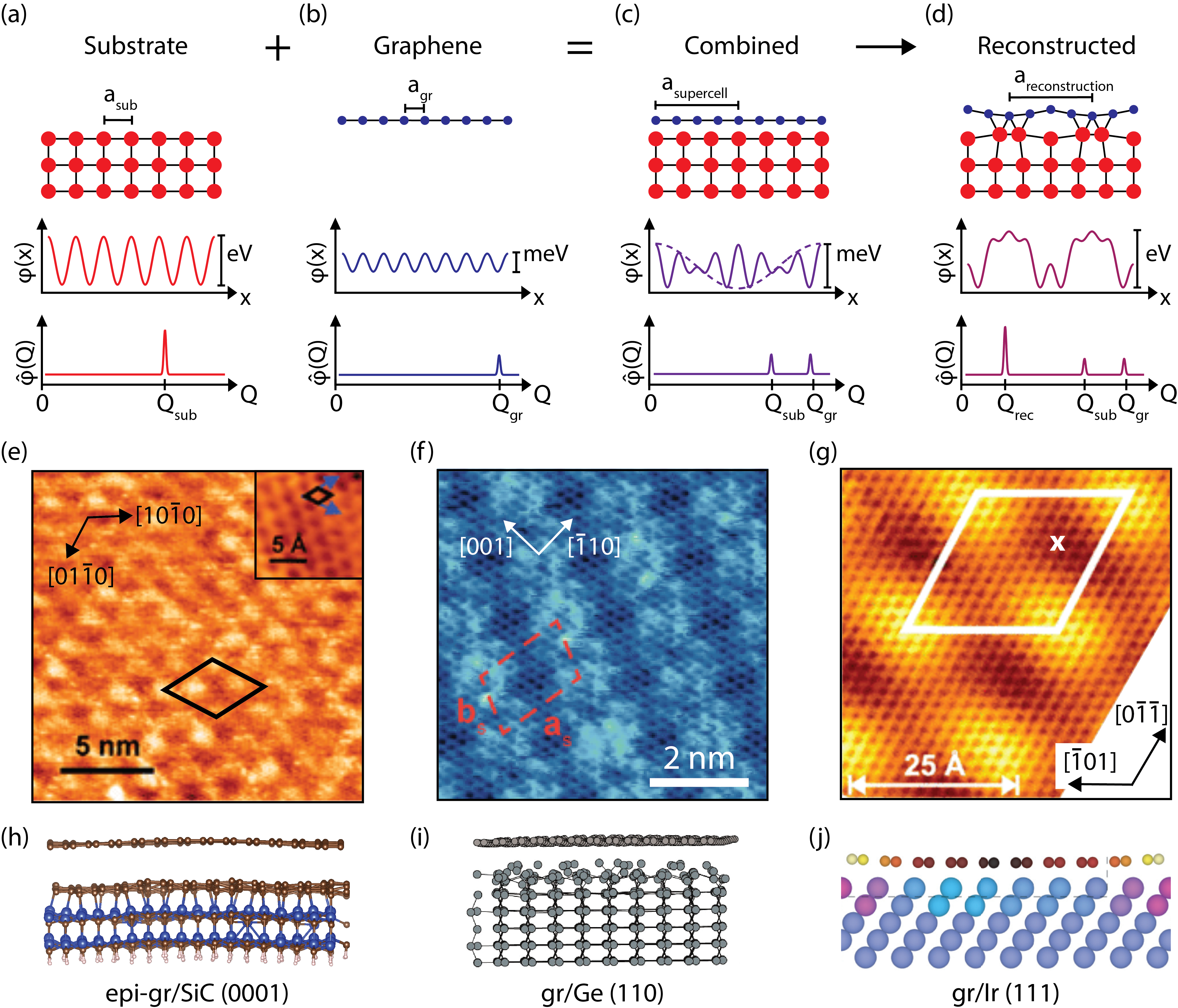}
    \caption{\textbf{Graphene induced surface reconstructions.} (a-d) Schematic lattice potential $\varphi(x)$ above crystalline surfaces and their Fourier transform $\hat{\varphi}(Q)$. (e) STM image of the $(6\sqrt 3 \times6\sqrt 3)$ $R30\degree$ reconstruction of buffer-layer graphene/SiC(0001) beneath the first epitaxial layer of graphene, inset shows the lattice of epitaxial graphene. Adapted from Poon \textit{et. al.} Phys. Chem. Chem. Phys., 12, 13522-13533 (2010) \cite{CoGrSiC_Sticking} with permission from the Royal Society of Chemistry. (f) STM image of the ``$6\times2$'' reconstruction of graphene/Ge(110). From Campbell \textit{et. al.} Phys. Rev. Materials 2, 044004 (2018) \cite{cambell2018_GrGe6x2} with permission from the American Physical Society. (g) STM image of the graphene/Ir (111) with $(10 \times 10)_{gr} / (9 \times 9)_{Ir}$ moiré structure. The x marks the HCP site that is favored for nucleation. From N'Diaye \textit{et. al.} Phys. Rev. Lett. 97, 215501 (2006) \cite{NDiaya2006_IrGrIr} with permission from the American Physical Society. (h,i,j) Cross-sectional structural models of graphene on SiC(0001), Ge(110), and Ir(111). Graphene/Ge model from Rogge \textit{et. al. }MRS Commun. 5, 539–546 (2015) \cite{Rogge2015_Ge110_6x2_Model} under Creative Commons Attribution 4.0 International License. Graphene/Ir adapted from Hämäläinen \textit{et. al.} Phys. Rev. B 88, 201406(R) (2013) \cite{Hamalainen2013_RrIr_Moire_Image} with permission from the American Physical Society.}
    \label{fig:reconstruction}
\end{figure*}

Beyond the simple-out-of plane relaxation models of Figs. \ref{polarity} and \ref{interference}, in many clean systems, strong graphene/substrate interactions induce reconstructions within the plane, in addition to the buckling relaxations out-of-plane. These reconstructions introduce a new periodicity $a_{reconstruction}$ which can be analyzed with the same Fourier analysis (Fig. \ref{fig:reconstruction}(d)) and challenge the idea of graphene being ``transparent'' with regard to the substrate. We highlight a few representative examples and their impacts on epitaxial film ordering, drawing on the broader field of moiré epitaxy.

Buffer graphene on Si-face SiC (0001), which forms a $(6\sqrt 3 \times6\sqrt 3)R30 \degree$ reconstruction, is a canonical example. A scanning tunneling microscopy (STM) image of this surface is shown in Fig. \ref{fig:reconstruction}(e, h). Here the buffer carbon layer is not fully graphitic. Instead, core level spectroscopy reveals that it has mixed hybridization with $\sim75\%$ $sp^2$ and $\sim25\%$ $sp^3$ character \cite{GrSiC_Bonding_XRSW}, strong bonding to the SiC substrate, out-of-plane dangling bonds, $0.4$ \AA\ buckling in the graphene, and both in-plane and out-of-plane distortions in the first layer of SiC \cite{GrSiC_Bonding_XRSW, emtsev2008interaction, van1975leed, forbeaux1998heteroepitaxial}. This deviation from graphitic $sp^2$ character also opens a bandgap in buffer graphene \cite{n2017band, nevius2015semiconducting}. The bandgap is important because it reduces the free carrier screening of the substrate potential, compared to the screening from metallic graphene. Thus, the remote potential through a semiconducting buffer graphene is expected to be larger than the remote substrate potential through metallic graphene. However, the buckled buffer layer, with its dangling bonds, may introduce larger local potential fluctuations than flat graphene, potentially dominating over any remote field effects from the substrate.

\begin{figure}
    \centering
     \includegraphics[width=1\linewidth]{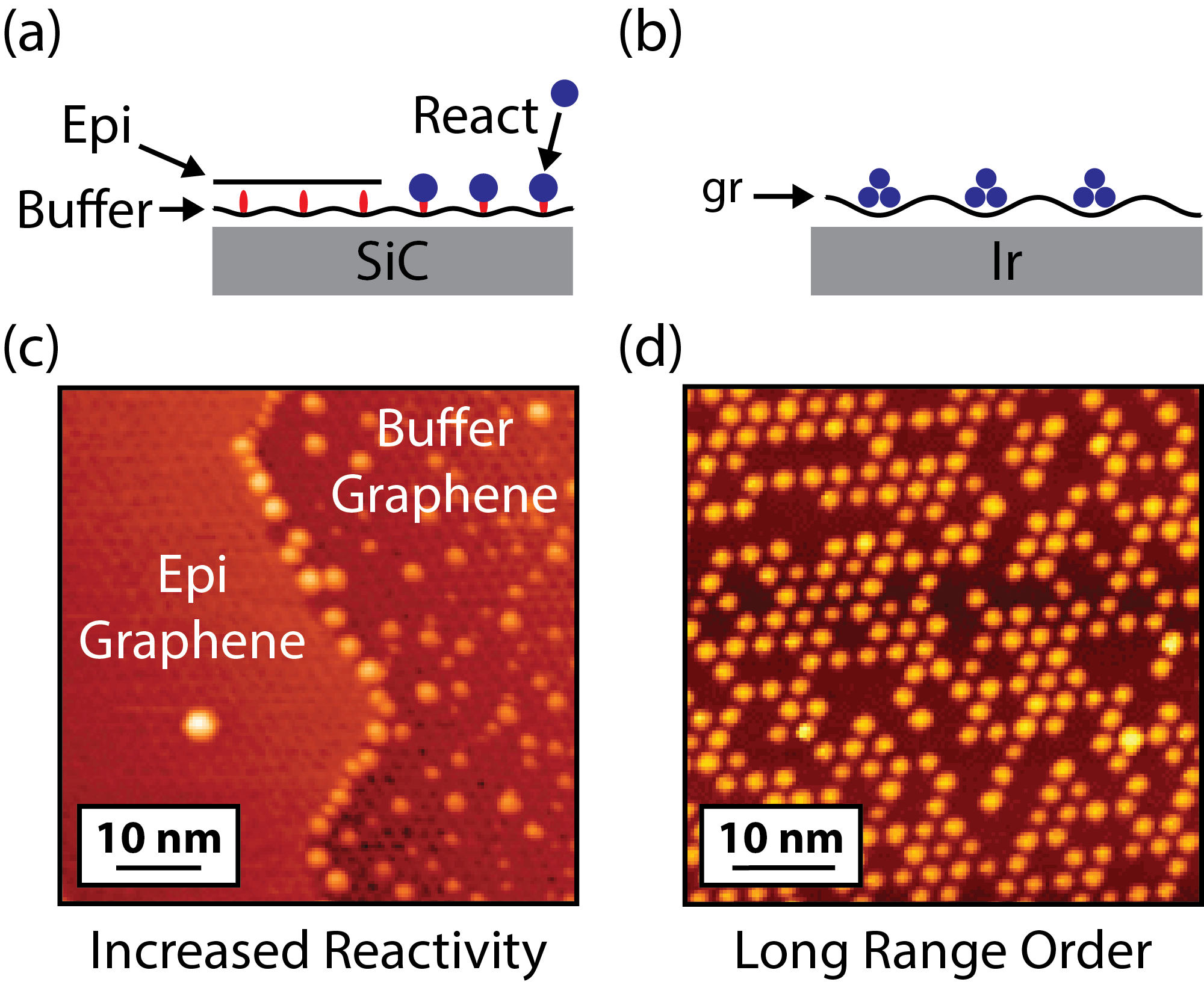}
    \caption{\textbf{Impact of graphene moiré and reconstructions on nucleation and epitaxy.} (a,b) Models for adatom sticking and ordering on graphene/SiC and graphene/Ir. (c) STM image of preferred Co sticking on buffer graphene vs. epitaxial graphene on SiC(0001). Adapted from
    Poon \textit{et. al.} Phys. Chem. Chem. Phys., 12, 13522-13533 (2010) \cite{CoGrSiC_Sticking} with permission from the Royal Society of Chemistry. 
    (d) Ordered superlattice of Ir clusters that follows the moiré peridicity of graphene/Ir(111). Adapted from N'Diaye \textit{et. al.} Phys. Rev. Lett. 97, 215501 (2006) \cite{NDiaya2006_IrGrIr} with permission from the American Physical Society.}
    \label{fig:dep}
\end{figure}

The large structural distortion of buffer graphene has a significant impact on adatom sticking and film growth. Figure \ref{fig:dep}(c) shows an STM image of Co atoms deposited on a surface that contains regions of buffer graphene and regions with a second ``epitaxial'' layer of graphene on top of the buffer \cite{CoGrSiC_Sticking}. Co clusters stick with higher probability to buffer graphene, than to epitaxial graphene. This preferred sticking on buffer graphene likely arises from the dangling bonds in $sp^2/sp^3$ hybridized buffer graphene, compared to the more idealized $sp^2$ character of epitaxial graphene. Presumably this change in bonding character would be relevant for the growth of remote epitaxial films on SiC buffer layer, indicating a large role for reconstruction in these systems.

Graphene on Ge (110) also reconstructs, into an apparent ``$6 \times 2$'' pattern (Fig. \ref{fig:reconstruction}(f, i)). This reconstruction is characterized by lateral and out of plane reconstructions of the top two Ge layers, with minimal distortions in the graphene itself \cite{cambell2018_GrGe6x2, chen2020_GrGe110_6x2, Kiraly2015_GrGe}. Here the graphene retains its semimetallic character. Note while this reconstruction is commonly referred to as ``$6 \times 2$'', its true periodicity is defined by $\mathbf{b} = \begin{bmatrix}
5 & \bar 1\\
2 & 2
\end{bmatrix} \mathbf{a}$, where $\mathbf{b}$ are the vectors of the surface reconstruction and $\mathbf{a}$ are the $(1\times1)$ Ge(110) surface unit cell vectors \cite{cambell2018_GrGe6x2}.

Finally, even graphene on metallic surfaces induces reconstructions. Fig. \ref{fig:reconstruction}(g, j) shows a STM image of graphene on Ir (111). The $\sim$10\% lattice mismatch between graphene and Ir produces a supercell with $\sim 2.5$ nm periodicity, corresponding to $(9 \times 9)$ Ir unit cells and $(10 \times 10)$ graphene unit cells. Photoemission and scanning probe measurements reveal that the graphene is buckled by $\Delta z \approx 0.3$ \AA, due to the periodic modulation in Ir-C atomic alignment, which changes the Ir $5d$ - C $2p_z$ hybridization strength \cite{dedkov2015SPM_GrMetals, Preobrajenski2008_GrMetals_XPS, NDiaya2006_IrGrIr}. 
Moiré-driven reconstructions have also been observed on gr/Ru(0001)\cite{GrRuSXRD,GrRu_STM_FigImage,Borca2010_GrRuMoire}, gr/Rh(111)\cite{HOLTSCH2018_GrRh_moire,Sicot2010_NiGrRh_nucleation}, gr/Ni(111)\cite{ZOU2018_GrNi_Moire,dahal2014_GrNi_Review}, gr/Pt(111)\cite{Martínez2016_GrPt_moire}, and gr/Cu(111)\cite{SULE2014_grCu_moire,soy2015_PtGrCu,gao2010_GrCu_Moire} with interaction strengths spanning from strong chemisorption (e.g., Ni, Rh) to weak van der Waals bonding (e.g., Cu, Pt) \cite{Preobrajenski2008_GrMetals_XPS, dedkov2015SPM_GrMetals}.

These graphene/metal moiré lattices are known to template the ordered assembly of adatoms and adatom clusters. Figure \ref{fig:dep}(d) shows an STM image of a fractional monolayer of Ir deposited on graphene/Ir(111). The Ir crystallizes in ordered arrays of metal clusters that follow the $(9 \times 9)_{Ir}/(10\times 10)_{gr}$ moiré periodicity \cite{NDiaya2006_IrGrIr}. Here, the Ir clusters preferentially nucleate at the hexagonal close packed (HCP) sites within the moiré cell. Similar moire ordered assembly has been observed in other graphene/metal systems including Pt deposition on graphene/Ru(0001) \cite{Pan2009_PtGrRu,Donner2009_PtGrRu,Sicot2010_NiGrRh_nucleation, Semidey2013_RhAuGrRu_DFT}, Pt deposition on graphene/Cu(111)\cite{soy2015_PtGrCu}, and Ni deposition on graphene/Rh(111)\cite{Sicot2010_NiGrRh_nucleation}.

Epitaxial ordering on graphene/metal moiré templates lies in apparent contrast to the polar bonding hypothesis for remote epitaxy. Recall that the electrostatic potential analysis by Kong \textit{et. al.} suggested the remote potential fluctuations should be strongest for materials with polar bonding \cite{kong2018polarity}. In contrast, Ir (111) has metallic bonding. For Ir on graphene/Ir (111) the ordering is thought to be driven by the reactivity of Ir to the HCP sites within the moiré cell \cite{NDiaya2006_IrGrIr}, rather than remote epitaxy through transparent graphene.

The experiments on moire epitaxy raise the question of which lattice would the film be interacting with: the bulk substrate or the graphene induced reconstruction (or both)? While the above experiments clearly demonstrate that long-wavelength (few nanometer) moire potentials can template ordered arrays of adatom clusters, it is less clear how that long wavelength periodicity might control crystallographic orientations of films or clusters, since the characteristic spacing between atoms is few Angstrom. From the perspective of nearest-neighbor ordering in a growing crystal, several nanometer wavelength moire potentials may be too long to dictate crystallographic ordering. On the other hand, for intermediate size reconstructions, where the reconstruction wavelength is just a few multiples of the native crystal lattice constant, it is plausible that these reconstructions could be important to determining crystallographic orientation, especially if epitaxy occurs via a moderate wavelength coincident site lattice relationship rather than a simple 1:1 film:substrate relationship. Note that local variations in graphene/substrate stacking order within a moire cell can control the orientation of small molecules \cite{GrRu_STM_FigImage}, suggesting that when the characteristic length scale for orientational ordering ($\sim$ nm for molecules) matches the wavelength of the potential fluctuation, then reconstructions or moires can indeed induce that type of ordering.

\section{How do graphene/substrate interactions tune adatom kinetics?} \label{sec:kinetics}

Just as the underlying substrate is expected to tune the remote lattice potential, the substrate is also expected to tune adatom diffusion on graphene-covered surfaces. Growth kinetics have been underexplored in the context of remote epitaxy, but have been considered more extensively in the surface science literature.Both the presence of graphene and the identity of the underlying substrate have significant impacts on the adsorption energies, sticking coefficients, and surface diffusion of adatoms on graphene-covered surfaces. Regarding adsorption and sticking, Manzo et al.,\cite{manzo2023nucleation} showed that the sticking of Ga and As on graphene/Ge (111) is significantly reduced compared to sticking on bare Ge(111). Poon \textit{et. al.}\cite{CoGrSiC_Sticking} showed that Co adatom clusters stick with higher probability to buffer graphene than to epitaxial graphene on SiC, where buffer graphene is more reactive due to is more $sp^3$-like character compared to $sp^2$-like character of epitaxial graphene (Fig. \ref{fig:dep}(c)).

\begin{figure}
    \centering
    \includegraphics[width=1\linewidth]{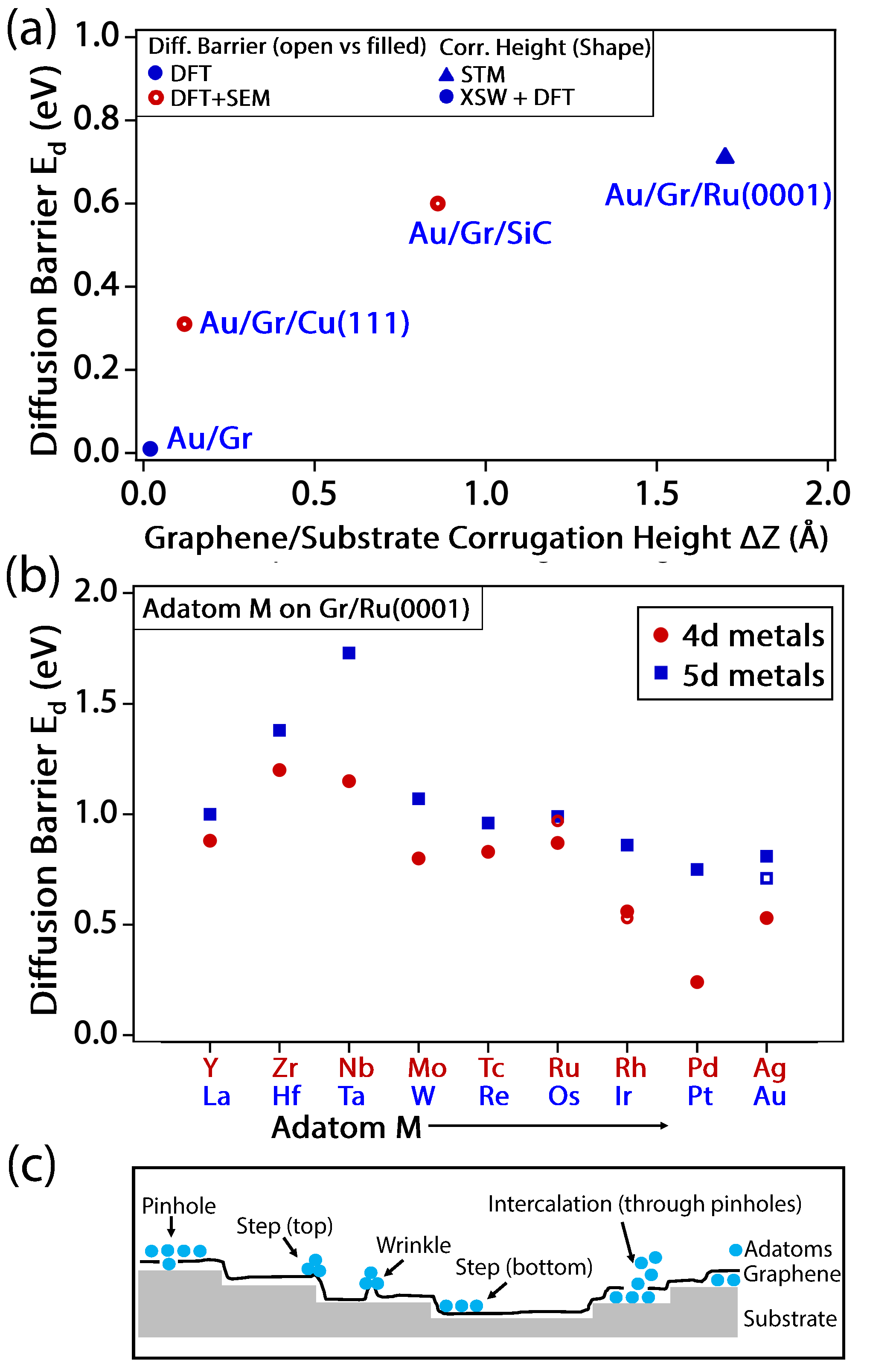}
    \caption{\textbf{Tunable surface diffusion barriers on graphene-covered substrates.} (a) Diffusion barrier for an Au adatom on graphene/substrate, for different substrates. (b) Diffusion barrier for different adatoms on graphene/Ru (0001).  DFT computed diffusion barrier data from Refs \cite{Amft_2011,JENSEN2004173,PhysRevB.58.13874_Au_HOPG,LIU2015397, 10.1063/1.4803893_Au_Rh_Ru, PhysRevLett.97.215501_PdPtIrTa_Ru0001, 10.1063/1.4934349_Ru_Ru0001,10.1063/1.4866876_Au_Gr_Cu111}. Measured/computed corrugation heights, measured by STM and by X-ray standing wave (XSW) are from Refs \cite{PhysRevLett.132.196201_Gr_Cu111, 10.1063/1.3309671_Ru_Pt_cor, PhysRevLett.101.126102, PhysRevLett.114.106804_Gr_SiC}.
    (c) Typical defects observed on graphene-covered surfaces. Defects are expected to be the nucleation sites for remote epitaxy and for competing mechanisms. }
    \label{fig:Diffusion barrier V2}
\end{figure}

Fig. \ref{fig:Diffusion barrier V2} summarizes the DFT computed surface diffusion barrier heights $E_d$ for several metal adatoms on graphene-covered metal surfaces \cite{Amft_2011,JENSEN2004173,PhysRevB.58.13874_Au_HOPG,LIU2015397, 10.1063/1.4803893_Au_Rh_Ru, PhysRevLett.97.215501_PdPtIrTa_Ru0001, 10.1063/1.4934349_Ru_Ru0001,10.1063/1.4866876_Au_Gr_Cu111}. In Fig. \ref{fig:Diffusion barrier V2}(a) we plot $E_d$ versus the measured or computed topographical corrugation height of graphene on these surfaces, since this corrugation is a proxy for the graphene-substrate interaction strength \cite{Preobrajenski2008_GrMetals_XPS}. As the substrate is changed, $E_d$ for Au adatoms systematically increases going from bare graphene ($\Delta z=0.02$ \AA)\cite{PhysRevLett.114.106804_Gr_SiC} to graphene/Cu (111) ($\Delta z = 0.12$ \AA),\cite{PhysRevLett.132.196201_Gr_Cu111} to buffer graphene/SiC ($\Delta z = 0.86$ \AA)\cite{PhysRevLett.114.106804_Gr_SiC} to graphene/Ru(0001) ($\Delta z = 1.7$ \AA).\cite{10.1063/1.3309671_Ru_Pt_cor} For a fixed graphene/Ru(0001) substrate and varying the adatom (Fig. \ref{fig:Diffusion barrier V2}(b)), the barrier height also changes significantly and follows a general trend of valence electron count of the adatom. These data suggest that kinetics, which are tunable via the graphene/substrate interface, may play a significant role in tuning the growth mechanisms and structure of films grown on graphene-covered surfaces, in addition to the thermodynamic conditions discussed in the previous sections. ``Remote'' tuned diffusion barriers go beyond the typical kinetic tuning knobs of growth rate and sample temperature.

These kinetic factors are essential, yet underexplored, parameters for tuning between the anticipated remote epitaxy and other competing growth mechanisms. For conventional epitaxy on covalent surfaces, the growth mode is highly dependent on the relations between surface diffusivity $D$, adatom flux $j$, and spacing between atomic steps on the substrate $L_{step}$. Note that $D$ relates to the barrier height via $D =D_0 exp(E_d/k_B T)$.
From these parameters one can define a dimensionless Péclet number $Pe = (L_{step}^2 \cdot j)/D$ that separates different growth modes: $Pe<1$ leads to step-flow growth, $Pe>1$ to layer-by-layer growth, and $Pe>>1$ to statistical growth \cite{tsao2012materials}.

For epitaxy on graphene-covered surfaces, the kinetic regimes are not well known, although one limiting case has been studied. For transferred graphene-covered surfaces, where pinholes are the primary defect, Manzo \textit{et. al.} showed that when the diffusion length $\lambda$ is greater than the spacing between pinholes $L_p$, nucleation occurs preferentially at pinholes where there is direct bonding to the substrate, rather than a true ``remote'' epitaxy mechanism \cite{manzo2022pinhole}. Note that $\lambda = 2\sqrt{D\tau}$ where $\tau$ is the diffusion time. This selective nucleation is followed by lateral coalescence of a continuous film. Since the pinholes are typically only a few nanometers in diameter, films grown by the pinhole mechanism can be exfoliated, and thus share many of the macroscopic hallmarks that are expected for films that grow by a remote epitaxy mechanism. Further studies are required to investigate how tuning surface diffusion, and tuning both the identity and concentration of different types of defects, tunes the growth mode on graphene-covered surfaces. While a pinhole mechanism dominates when there is a high concentration of pinholes or larger openings in graphene \cite{manzo2023nucleation, lim2022selective, jeong2020selective, kim2022graphene, manzo2022pinhole, fazlioglu2024atomic, zulqurnain2022defect}, perhaps other sites such as graphene-covered atomic step edges or local variations in the graphene/substrate registry could be the nucleation sites for a true remote epitaxy mechanism (Fig. \ref{fig:Diffusion barrier V2}(c)).

\section{Comparison with experiment}\label{sec:experiment}

\begin{figure*}
    \centering
    \includegraphics[width=0.9\linewidth]{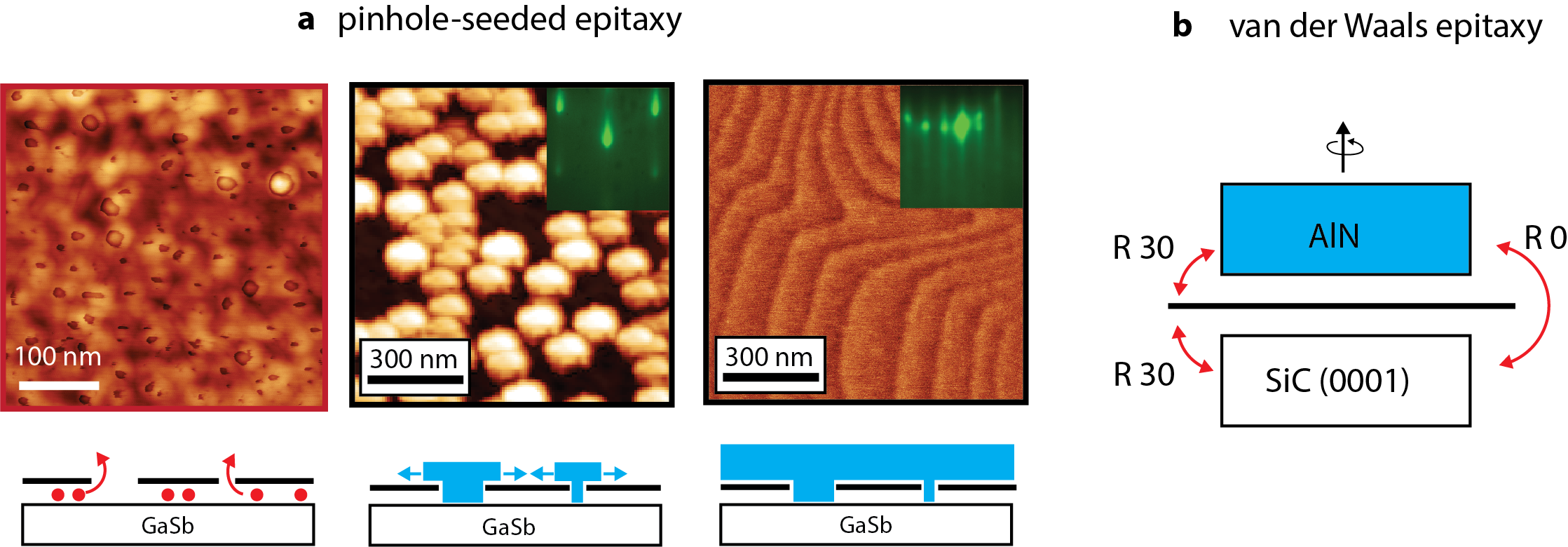}
    \caption{Challenges for distinguishing remote epitaxy from competing growth mechanisms. (a) Pinhole seeded epitaxy of GaSb on transferred graphene/GaSb (0001), from Ref. \cite{manzo2022pinhole}. Unintentional pinholes are created from desorption of trapped interfacial oxides (red circles). Adapted from Manzo, \textit{et. al.} Nature Communications, 13, 4014 (2022) \cite{manzo2022pinhole} under Creative Commons CC BY.
    (b) Serial van der Waals epitaxy. Since epitaxial graphene grows with a 30 degree in plane rotation compared to SiC, and AlN grows with a 30 degree rotation compared to graphene, it is difficult to distinguish this series of vdW epitaxial alignment to graphene, versus true ``remote'' interactions through graphene.}
    \label{fig:pinhole_vdw}
\end{figure*}

Our analysis of the literature suggests that the remote substrate potential is not as strong as often assumed. Rather, its typical magnitude (tens of meV) is similar to the direct potential from graphene and the potential from graphene-induced interfacial reconstructions. What explains the existing experiments on remote epitaxy?

We argue there are two competing mechanisms, pinhole-seeded epitaxy \cite{manzo2022pinhole} and van der Waals epitaxy to graphene\cite{laduca2024cold}, that compete with remote epitaxy and are often difficult to distinguish from a true remote mechanism. Pinhole epitaxy often occurs because defects are strong ($\sim$ eV) local perturbations to Eq 2 compared to the native meV fluctuations above clean graphene (Fig. \ref{fig:reconstruction}). Most transferred graphene platforms are not clean enough \cite{manzo2022pinhole, kim2021impact,kim2021role}. VdW epitaxy to graphene occurs in systems with weak remote substrate potential, where lattice matching to graphene competes \cite{laduca2024cold}. It is difficult to distinguish from ``remote'' epitaxy if there are serial epitaxial relationships between film and graphene, and graphene and substrate. We briefly survey the main experimental observations, explain how many of these observations can be explained by these alternate mechanisms, and conclude with one experiment that cannot be explained by pinholes or epitaxy to graphene \cite{du2022controlling}.

\textbf{Lack of observed pinholes by microscopy.} A lack of observed pinholes in graphene, either before\cite{kong2018polarity,kim2017remote} or after\cite{chang2023remote} epitaxial growth, is often invoked as evidence to rule out pinhole seeded lateral epitaxy, in favor of remote epitaxy. The challenge with the before-growth observations is that they typically overlook a critical step: annealing the graphene/substrate interface. 
Previous experiments\cite{manzo2022pinhole, kim2021impact} demonstrate that the standard wet and semi-dry graphene transfer procedures leave trapped contaminants and native oxides at transferred graphene/III-V substrate interfaces. Upon annealing to temperatures required for III-V film epitaxy, these contaminants volatilize to create pinholes in the graphene with densities of $10^3$ $\mu$m$^{-2}$, three orders of magnitude over thresholds needed to observe true intrinsic remote epitaxy \cite{manzo2022pinhole} (Fig. \ref{fig:pinhole_vdw}a). Most graphene characterization for remote epitaxy is performed this critical annealing step, and thus cannot rule out pinholes. Pinholes down to the substrate often serve as the reactive sites for selective area epitaxy, followed by lateral coalescence of continuous films.

The challenge with post-growth defect analysis is that buried defects in graphene are difficult to resolve. Moreover, sufficient statistics are required to rule out their impact on the growth mechanism. The lack of an observed graphene pinhole by cross sectional STEM of an individual BaTiO$_3$ island grown on graphene/SrTiO$_3$ (001) was cited as evidence favoring remote epitaxy \cite{chang2023remote}. However, greater statistics and larger field of view imaging are required to understand if the alignment of the BaTiO$_3$ island was due to ``remote'' interactions, random polydispersity, or due to features outside the field of view of individual images. Cross sectional TEM is inherently challenging for such analyses, since one of the surface directions is viewed in projection, through a finite thicknesses (typically $\sim 50-100$ nm) along the electron beam direction. Sophisticated techniques like ptychography are required to resolve such depth dependent information. These advanced techniques, which inherently focus on small fields of view, make it inherently difficult to generate sufficient statistics.

\textbf{Epitaxy ``through'' multiple layers of graphene.} Epitaxial films have been reported for growth on few-layer graphene on III-V and oxide substrates \cite{kong2018polarity}. For these systems it is assumed that pinholes are less likely to dominate the growth mechanisms on multilayer graphene, since such pinholes would need to align vertically. However, if pinholes are created by desorption of native oxides or contaminants at the few layer graphene / substrate interface, these pinholes would be aligned \cite{manzo2022pinhole}. Moreover, the significant attenuation due to screening suggests that epitaxy through more than two or three graphene layers is unlikely \cite{kawasaki2025model}.


\textbf{Ability to exfoliate a membrane.} The ability to exfoliate a membrane from a (graphene-covered) substrate is often invoked as evidence for remote epitaxy \cite{kim2017remote, kong2018polarity,han2022remote,liu2025remote}. However, films seeded at pinholes can also be exfoliated \cite{manzo2022pinhole}. Moreover, metal stressor layers deposited on top of the membrane are also commonly used to aid in exfoliation: presumably such stressors would not be needed if the film / graphene / substrate interfaces have sufficiently low defect density and have primarily van der Waals interactions. Finally, the presence of a graphene interlayer is not strictly required for exfoliation: Zhang et. al. \cite{zhang2025atomic} report efficient spalling of epitaxial oxide membranes from oxide substrates, by selecting materials combinations with weak interfacial bonding.

\textbf{Film aligned to substrate on cleaner epitaxial graphene.} This argument is often invoked for III-nitride epitaxy on graphene / SiC (0001) \cite{wang2023modulation,lee2024gan,qiao2021graphene}. Here, the graphene is grown epitaxially on SiC (0001), and thus has lower defect and contaminant density than transferred graphene platforms. The challenge is that it is difficult to distinguish ``remote'' epitaxy from serial van der Waals epitaxy. Graphene ($a_{gr}=2.46$ \AA) grows epitaxially on SiC ($a_{SiC}=3.09$ \AA) in a commensurate site lattice condition where a $(13\times 13)$ graphene superstructure nearly matches a $(6\sqrt{3}\times6\sqrt{3})R30\degree$ supercell of SiC. AlN ($a_{AlN}=3.11$ \AA) has nearly identical lattice parameter as SiC, and thus a similar $30\degree$ rotated commensurate site lattice relationship is expected between AlN and graphene. Thus for an all epitaxial AlN/graphene/SiC (0001) heterostructure, it is difficult to distinguish whether the structure forms via ``remote'' AlN-SiC interactions through graphene, or via a series of direct epitaxial interactions between AlN / graphene, and then graphene / SiC (Fig. \ref{fig:pinhole_vdw}b). Both mechanisms would produce the same oriented structure.

\begin{figure}
    \centering
    \includegraphics[width=0.95\linewidth]{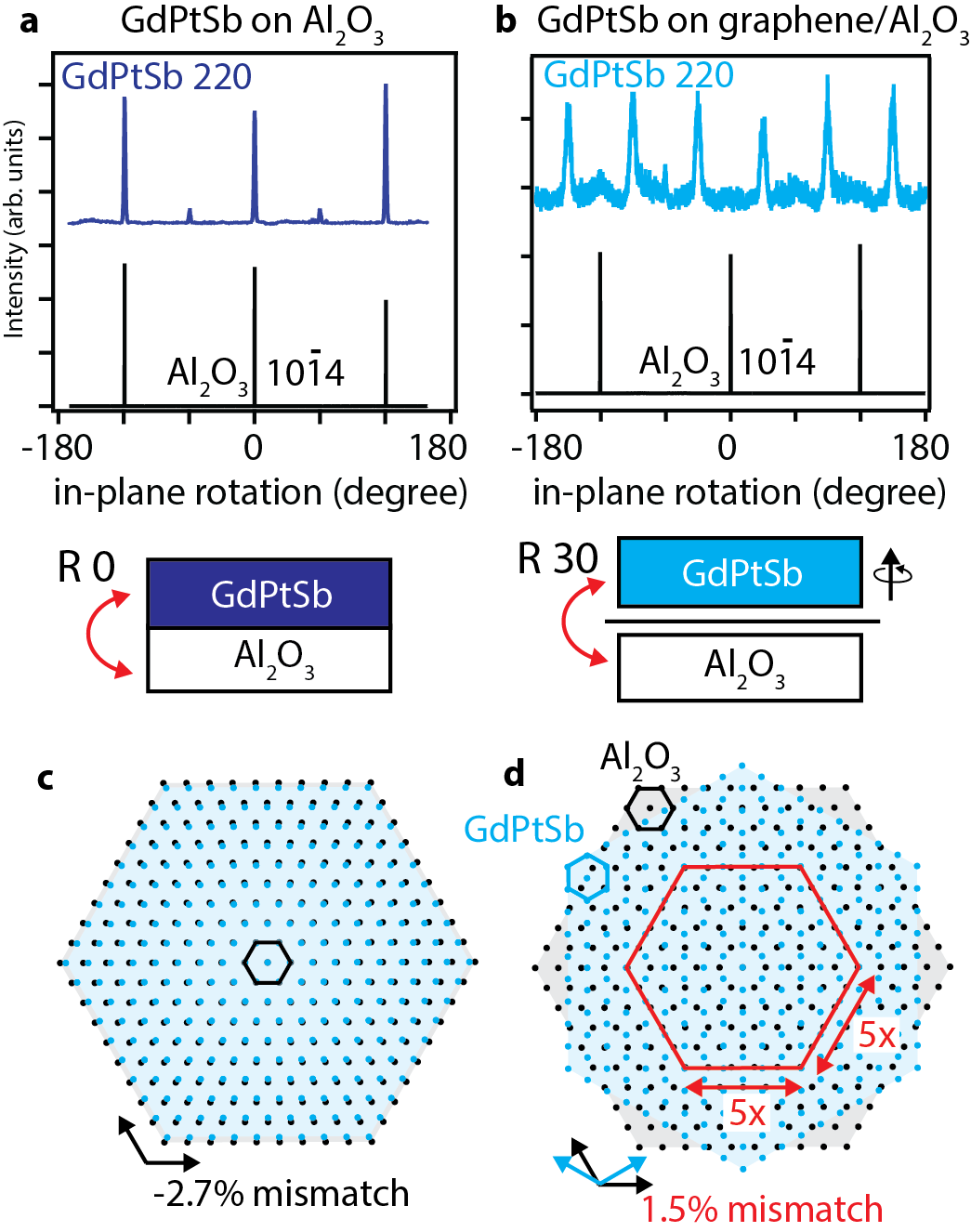}
    \caption{Long-range evidence suggestive of remote epitaxy, that cannot be explained by pinholes or van der Waals epitaxy. (a) Azimuthal x-ray diffraction shows GdPtSb grows on c-plane sapphire with a simple hexagon-on-hexagon R30 epitaxial orientation. (b) GdPtSb grows on polycrystalline graphene-covered sapphire with a long-range 30 degree in plane rotation compared to direct growth on sapphire. (c,d) Corresponding in-plane lattices for sapphire and GdPtSb. The red $(5\times 5)$ supercell has a smaller effective lattice mismatch than the $(1\times 1)$ cell. Adapted from E. Du, \textit{et. al.} Nano Letters, 22, 21 (2022) \cite{du2022controlling}, with permission from the American Chemical Society.}
    \label{fig:R30}
\end{figure}

\textbf{Excluding pinholes and mutual vdW epitaxy: rotated GdPtSb growth on graphene/sapphire.} We end with one observation that, to our knowledge, cannot be explained by pinholes or vdW epitaxy.
Du et. al.\cite{du2022controlling} found that GdPtSb films grow on polycrystalline graphene / Al$_2$O$_3$ (0001) with a $30\degree$ rotation compared to direct epitaxy of GdPtSb on sapphire (Fig. \ref{fig:R30}). This rotated ordering cannot be explained by pinholes (which produce a $0\degree$ orientation) or vdW epitaxy to the polycrystalline graphene (which would produce random in-plane oriented GdPtSb). Importantly, the new epitaxial orientation for GdPtSb on graphene/sapphire is a long-range signature that does not rely on local microscopy to exclude pinhole or vdW mechanisms.

\section{Outlook}\label{sec:outlook}

How transparent is graphene? Our critical analysis of DFT slab calculations \cite{dai2022highly, kong2018polarity} and analytical modeling \cite{kawasaki2025model} suggests that for weakly interacting graphene/substrate systems, the remote potential of the substrate is significantly attenuated by graphene. For many commonly used substrates, the remote substrate potential and the graphene potential have similar magnitude. We proposed a Fourier analysis to analyze the interference patterns between graphene and substrate in a bias-free way. 
The literature on graphene-induced reconstructions suggest that systems with strong graphene/substrate interactions are a significant deviation from the transparent limit. In these systems, deviations from ideal $sp^2$ bonding and variations in local registry of the graphene lattice on the substrate lattice can significantly impact sticking and template long range ordering of adatoms. These effects are well studied in the limit of a few monolayers of adatoms on highly interacting graphene, and further studies are required to connect these insights to continuous films on graphene.

Many outstanding questions remain. What type(s) of interactions are most important for the ordering in remote epitaxy: covalent hybridization, van der Waals interactions, electrostatics, or some combination of these interactions? How can we distinguish a purely ``remote'' mechanism through transparent graphene, versus a series of direct substrate-graphene plus graphene-film interactions? How do kinetics tune the growth mechanisms, and how does nucleation on graphene-covered surfaces proceed? We suggest the following directions for the field:

\begin{enumerate}
    \item \textbf{Direct measurements of the remote potential}, to test model predictions and distinguish effects of the remote potential versus growth kinetics. AFM-based force-distance spectroscopy \cite{ke1999quantity, hoffmann2001direct, giessibl2001imaging, allain2017color} and thermal  desorption spectroscopy \cite{king1975thermal, habenschaden1984evaluation, apker1948surface} are two possible methods to measure covalent and van der Waals contributions. Photoemission electron microscopy (PEEM) and low energy electron microscopy (LEEM) \cite{schmidt1998speleem, bethge1983photo, zakharov2012recent}, Kelvin probe microscopy (KPFM) \cite{panchal2013standardization, behn2021measuring}, and ballistic electron emission microscopy (BEEM) \cite{bell2016ballistic, prietsch1995ballistic} are suited for measuring electrostatic contributions.

    \item \textbf{Explicit pairwise potential calculations that consider both the identity of the film and the substrate.} Bonding is inherently a pairwise interaction, where changing the identity of one species in the pair changes that interaction. For example, Ga-Ga bonding is different than Ga-As bonding. However, with the exception of the kinetics literature \cite{Amft_2011,JENSEN2004173,PhysRevB.58.13874_Au_HOPG,LIU2015397, 10.1063/1.4803893_Au_Rh_Ru, PhysRevLett.97.215501_PdPtIrTa_Ru0001, 10.1063/1.4934349_Ru_Ru0001,10.1063/1.4866876_Au_Gr_Cu111}, most calculations of the remote potential to date\cite{kim2017remote,dai2022highly,jiang2019carrier} consider only the identity of the substrate, and do not explicitly consider the identity of the film. The film is only considered afterwards, e.g. via the adjustable scale factor $\gamma$ in Eq 1 \cite{kong2018polarity}. A true understanding of the remote potential requires explicit consideration of the identity of both film and substrate.

    \item \textbf{Systematic growth studies that vary the kinetics}. Growth temperature, growth rates, precursor type, and defect identity and density are all key parameters that may enable tuning of remote epitaxy versus other mechanisms \cite{manzo2022pinhole}. The question of whether remote epitaxy ``works'' for a particular film / graphene (or other 2D material) / substrate system may not have a simple binary answer. Rather, it should also depend on what pathways are kinetically accessible.

    \item \textbf{Cleaner graphene/substrate systems} that avoid the use of solvents or etchants during transfer, or new methods for graphene growth directly on the substrate of interest. Cleaner graphene is critical identify intrinsic contributions to remote epitaxy rather than other mechanisms like pinhole-seeded lateral overgrowth.
    
\end{enumerate}

To provide relevant information, we suggest that future first-principles theoretical calculations of remote epitaxy address these issues as well. Specifically, they must provide isolated slab calculations which include graphene layer(s) above the substrate, any relevant reconstructions of the graphene/substrate, and different substrate and film atoms, separated by enough vacuum that the top of the film does not interact with the bottom of the substrate. Ideally, these calculations would capture information relevant to the kinetics of growth such as diffusion barriers for individual atoms or sliding barriers for islands. While requiring significant additional computation, calculations which also include graphene pinholes, defects, and step edges would be invaluable for understanding how these impact real-world bonding potentials and growth.

\section{Acknowledgments}

JKK thanks Michael Arnold, Paul Evans, and Chris Palmstrom for helpful discussions.

This work was primarily supported by the U.S. Department of Energy, Office of Science, Basic Energy Sciences, under award no. DE-SC0023958 (AS and JKK). ZL and QTC were supported by the Laboratory Directed Research and Development Program at Sandia National Laboratories under project 233271. JKK and ZL acknowledge preliminary support from the Air Force Office of Scientific Research (FA9550-21-0127).

This work was performed, in part, at the Center for Integrated Nanotechnologies, an Office of Science User Facility operated for the U.S. Department of Energy (DOE) Office of Science.
Sandia National Laboratories is a multi-mission laboratory managed and operated by National Technology \& Engineering Solutions of Sandia, LLC (NTESS), a wholly owned subsidiary of Honeywell International Inc., for the U.S. Department of Energy’s National Nuclear Security Administration (DOE/NNSA) under contract DE-NA0003525. This written work is authored by an employee of NTESS. The employee, not NTESS, owns the right, title and interest in and to the written work and is responsible for its contents. Any subjective views or opinions that might be expressed in the written work do not necessarily represent the views of the U.S. Government. The publisher acknowledges that the U.S. Government retains a non-exclusive, paid-up, irrevocable, world-wide license to publish or reproduce the published form of this written work or allow others to do so, for U.S. Government purposes. The DOE will provide public access to results of federally sponsored research in accordance with the DOE Public Access Plan.

\bibliographystyle{apsrev}
\bibliography{ref}

\end{document}